\input lecproc.cmm
\input epsf.tex
\newbox\grsign \setbox\grsign=\hbox{$>$} \newdimen\grdimen \grdimen=\ht\grsign
\newbox\simlessbox \newbox\simgreatbox \newbox\simpropbox
\setbox\simgreatbox=\hbox{\raise.5ex\hbox{$>$}\llap
     {\lower.5ex\hbox{$\sim$}}}\ht1=\grdimen\dp1=0pt
\setbox\simlessbox=\hbox{\raise.5ex\hbox{$<$}\llap
     {\lower.5ex\hbox{$\sim$}}}\ht2=\grdimen\dp2=0pt
\setbox\simpropbox=\hbox{\raise.5ex\hbox{$\propto$}\llap
     {\lower.5ex\hbox{$\sim$}}}\ht2=\grdimen\dp2=0pt

\def\simless{\mathrel{\copy\simlessbox}}

\contribution{ Gamma-ray flux variability in the sample of EGRET blazars } 
\contributionrunning{ Variability of EGRET blazars }
\author{ Pawe\l\ Magdziarz@1, Rafa\l\ Moderski@2@3, Greg M. Madejski@4 } 
\authorrunning{ P. Magdziarz, R. Moderski, G. M. Madejski }
\address{ 
@1Astronomical Observatory, Jagiellonian University,\newline 
Orla 171, 30-244 Cracow, Poland 
@2UPR 176 du CNRS, DARC, Observatoire de Paris, Meudon, France 
@3N. Copernicus Astronomical Center, Bartycka 18, 00-716 Warsaw, Poland
@4LHEA, NASA/Goddard Space Flight Center, Greenbelt, MD 20771, USA 
}

\abstract{ 
We analyze average $\gamma$--ray variability statistics for the sample of
blazars detected by {\it CGRO}/EGRET. We re-reduce all the available EGRET
observations and analyze light curves by Monte Carlo modeling of the
variability statistics including observational artifacts. We show that
the observed variability behavior is dominated by the distribution of
measurement errors which leads to strong systematical effects in all of
the these statistics. General variability behavior detected by us is 
consistent with non-linear models and shows marginal correlation at long
time scales in the structure function. We determine limits on
distributions of the variability parameters with synthesis of flare 
population.  We conclude that this method shows that all blazar light
curves are consistent with a superposition of multiple flares.  
}

\titlea{1}{Introduction}

The EGRET archive provides important opportunity to study variability
behavior of the high energy Compton component of $\gamma$-ray blazars, which is
crucial for multi-wavelength cross-correlation analysis and for further
developing of theoretical models (e.g., Urry 1996; Madejski et al.\ 1996).
Although the EGRET archive was investigated extensively (e.g., McLaughlin
1996), the complex nature of the EGRET instrument (Thomson et al.\ 1993;
1995) together with a complex, non-stationary pattern of blazar
variability, makes conclusions uncertain. Effects of relativistic
Doppler beaming produce a broad range of time scales and amplitudes which
are strongly confused due to limited photon statistics and poor sampling.
Relatively short observations also prevent 
simple auto-correlation analysis. Moreover, statistical behavior of EGRET
flux measurements is still far from being understood (e.g., Mattox et al.\
1996), and confuses significantly the analysis. In order to understand the
real content of physical information, we re-analyze all publicly available
EGRET observations of blazars and model average variability
statistics including effects coming from complex distribution of
measurement errors. This paper presents preliminary results from our
analysis based on a method of flares population synthesis (Magdziarz \&
Machalski 1993). 

\titlea{2}{Data Reduction}

The results presented here are based on EGRET data in viewing
periods from 0.2 to 428 (1991 April 22 to 1995 September 20). We use all
events at E $>100$ MeV, and we get positions of sources from EGRET
catalogues (Thompson et al.\ 1995, 1996). The flux is calculated using
maximum likelihood method (Mattox et al.\ 1996) with the LIKE v5.0
software, and Galactic diffuse radiation is described using the model by
Hunter et al.\ (1996). The dominant contribution to the error of flux
measurements comes from fitting and subtraction procedure of the composite
background model which leads to complex, strongly non-linear flux-error
relation and makes the measurement error distributions hard to estimate
(Willis 1996).

\begfig 0 cm
\centerline{\epsfysize=5truecm\epsfbox{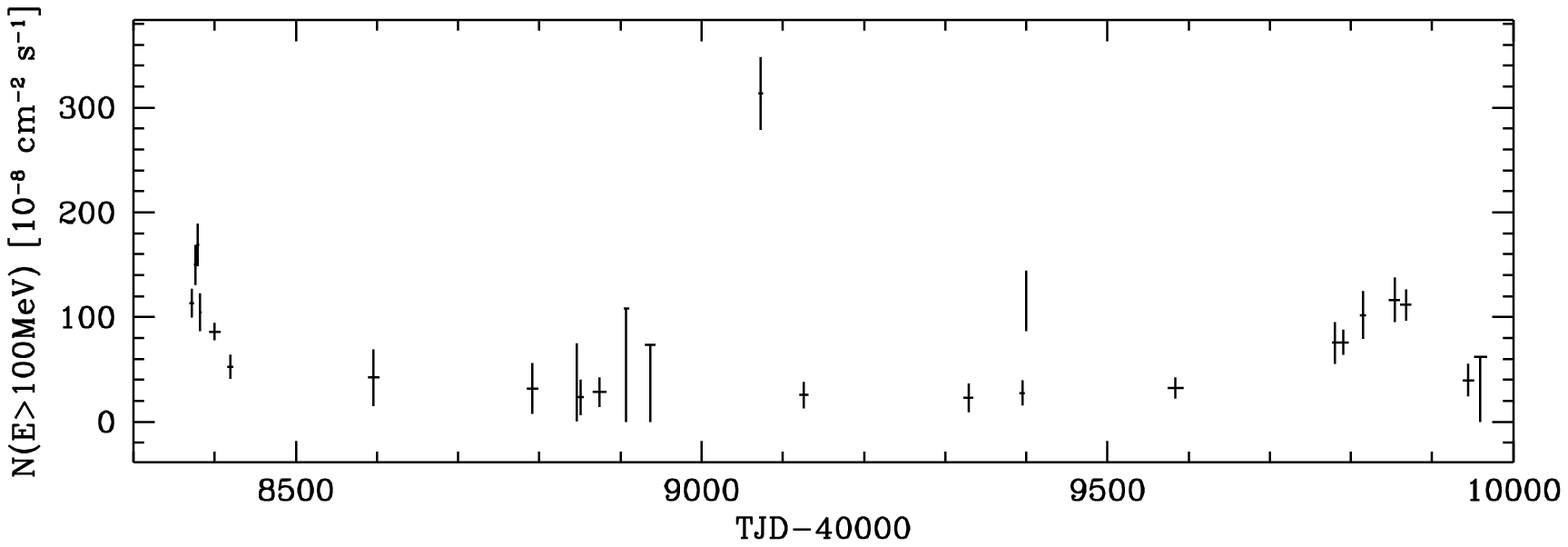}}
\figure{1}{
A light curve from EGRET observations of PKS~0528. Upper limits
mark detection below 1$\sigma$ level, error bars correspond to 95 per
cent confidence limit.
}
\endfig

\noindent
Fig.~1 shows an example of a light curve produced from EGRET observations of
a bright blazar PKS~0528. Although the baseline of emission seems to be
well observed in this particular source, sensitivity of the EGRET detector
(Thompson et al.\ 1993) is relatively low in respect to that necessary for
observation of complete light curves in most of the blazar sources.
Moreover, strong flaring behavior of the variability makes the light
curves dominated by observations near the detection limit. On average
about 60 per cent of observational measurements in the EGRET sample of 52
blazars give detection below 1$\sigma$ limit at the positions of sources known
from measurable flux in flaring epochs. The light curves also show also that
most of observations do not resolve flaring events in time, since an
average of 20 pointings per source is spread over 4 years of
observation. 

\titlea{3}{Variability Analysis}

Assuming a flaring behavior of EGRET blazars light curves, we construct a
simple phenomenological model of variability. The flares of radiation are
described by amplitude, duration time, and a repetition time lag. The
resulting light curve is a superposition of a population of flares (cf. 
Magdziarz \& Machalski 1993). Physics of the source determines some
distributions of the above phenomenological parameters which we constrain
from the data. In a such treatment, the non-linear behavior of variability
produces correlations between the parameter distributions. The instrument
sensitivity prevents an analysis of the flare duration, since the
data binned in time to a reasonable counts limit cannot resolve flares
appearing on time scales $\simless$ 1 day. This makes the first
order structure function degenerate, such that it is 
constant as characteristic for pure,
zero time scale, noise process (cf. Wagner 1996).  On the other hand, all
blazars in the EGRET sample show in their time history at least
some measurement epochs with apparent flux below sensitivity limit of the
instrument. This suggests that all of the EGRET measurements detect flare
events only, and any quiescent level of blazar emission is under detection
limit (cf.\ Hartman 1996; Willis 1996). Then the distribution of the flux
measurements corresponds directly to the distribution of flare amplitudes.
We investigate the variability characteristics as an average over the
whole sample since the data quality prevents conclusive analysis of the
most of individual sources. This treatment is supported by the evidence
that the emission mechanism in blazars is related exclusively to the
relativistic jet (e.g., Urry \& Padovani 1995), and all of the EGRET
blazars show significant variability when taking into account selection
effects by instrument detection limit (McLaughlin et al.\ 1996). 

\titleb {3.1}{Distribution of Apparent Amplitude}

Fig.~2 presents an average distribution of normalized flux measurements in
the sample. We normalize a light curve of each source to its average flux,
and next calculate by Monte Carlo method the distribution of fluxes
including measurement errors. The flux scale is renormalized to the
average over the sample. Comparison of the distributions for all of the
EGRET measurements (the solid line) and for measurements on the level
higher that 1$\sigma$ (the short-dashed line) shows homogeneity of the
sample up to the level of the detection limit. However, the distribution
simulated under assumption of constant internal flux of the source (the
dashed line) indicates that the overall behavior of the flux distribution
is dominated by measurement (i.e., statistical) errors. Deconvolution of
the internal flux distribution needs more sophisticated methods since the
reduction procedure produces a complex correlation between the measured
flux and its error, and the distribution of measurement errors is Gaussian
for high level detections only (Willis 1996). The apparent flux
distribution shows trace of internal variability up to the flux of $\sim
80\times 10^{-8}$ photons cm$^{-2}$ s$^{-1}$ where statistical errors
begin to dominate. This corresponds to the maximum apparent amplitude of
order $\sim 7$ (in the scale of average flux over the sample) which is, a
posteriori, consistent with the non-linear character of variability (e.g.,
McHardy 1996). A luminosity function for EGRET sample is poorly determined
(cf.\ Willis 1996 for discussion), and thus the conversion of physical
scales is uncertain making the flux distribution smeared out by space
distribution of sources.

\begdoublefig 0 cm
{\centerline{\epsfysize=6.3truecm\epsfbox{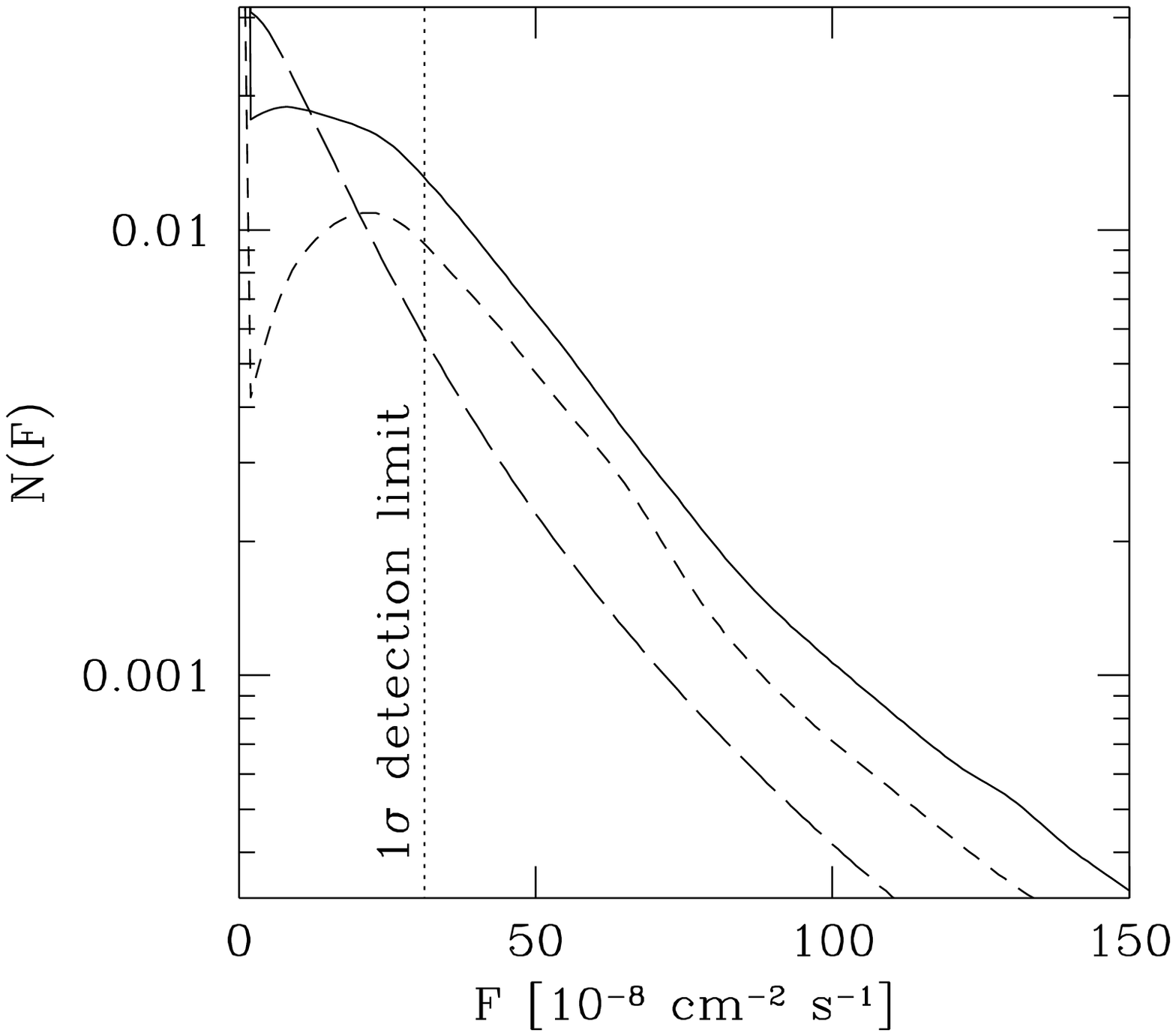}}
\figure{2}{
Distribution of normalized flux measurements averaged over the sample. 
The solid curve shows the distribution including all of the observations,
the short-dashed curve, result for detections better than 1$\sigma$, and
the long-dashed curve determines the distribution derived without internal
variability. 
}}
{\centerline{\epsfysize=6.3truecm\epsfbox{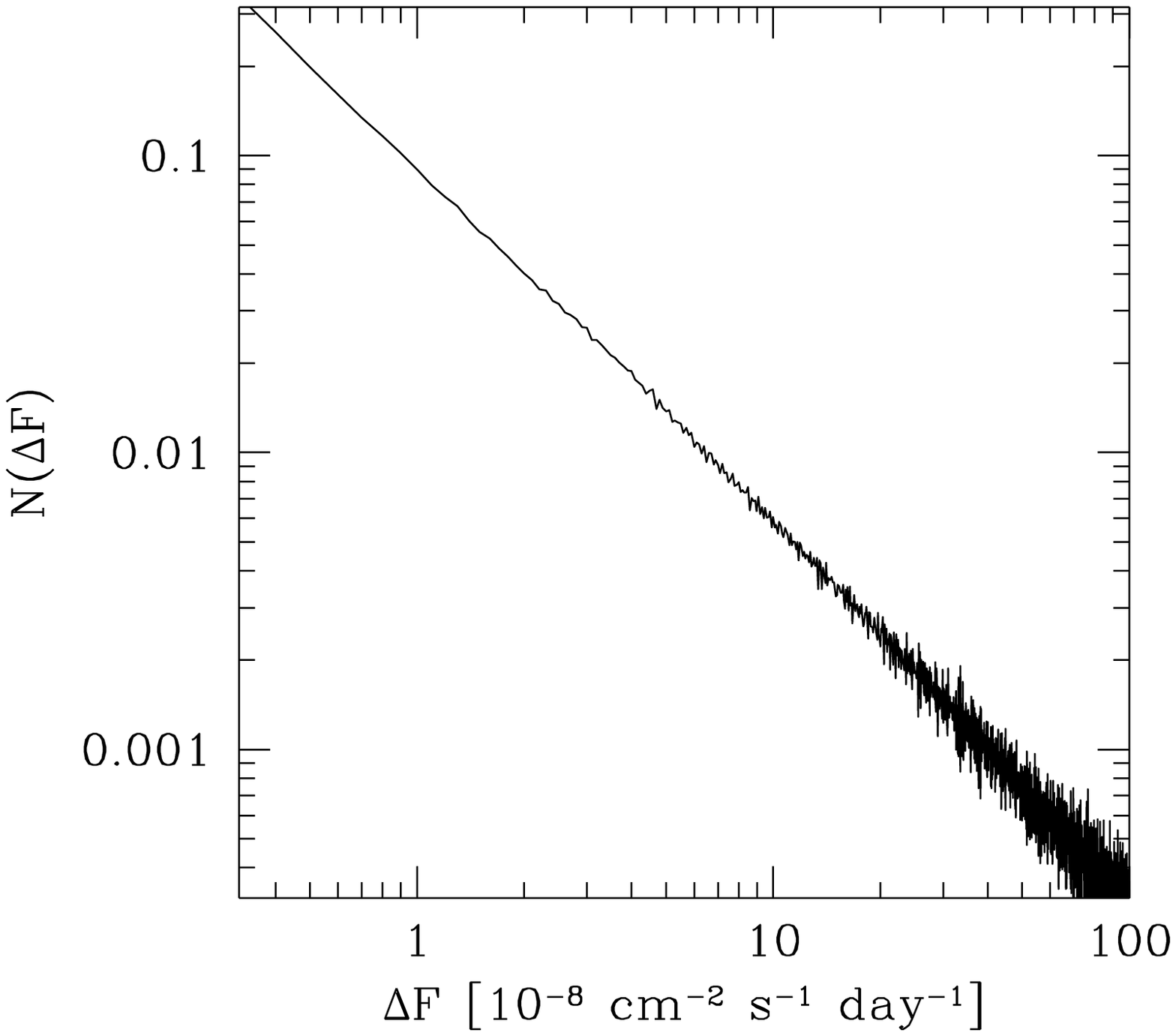}}
\figure{3}{
Distribution of normalized, apparent gradient averaged over the sample of
EGRET blazars. The distribution is homogeneous, and well described by a    
power law functional dependence. The gradient scale is renormalized to the
flux average over the sample.
}}
\enddoublefig

\titleb {3.2}{Distribution of Apparent Gradient}

The observed light curves in EGRET sample have in average 20 data points,
much less than necessary to study correlation decaying time scales (e.g.,
Sugihara \& May 1990). However, some information may be still retrieved
from an average distribution of flux temporal gradients which corresponds
to a distribution of the first derivative of the observed light curve.
Fig.~3 presents the average gradient distribution over the sample
calculated using the Monte Carlo method and including measurement errors.
The gradient scale is defined in respect to the average flux, just as for
the amplitude distribution. The distribution indicates no internal
structure and shows pure power-law functional dependence. The steep index
of the distribution indicates that small flares are much more frequent
than strong ones. However, despite of complex structure of light curves,
the variability process appears to be homogeneous up to the Nyquist
frequency limit of $\sim 0.07$ day$^{-1}$ defined by sampling. This is
generally consistent with the light curve produced by superposition of
identical flares of radiation with no indication for any duty cycles. The
average variability gradient estimated from the sample is $\sim 10^{-6}$
photons cm$^{-2}$ s$^{-1}$ day$^{-1}$, however, since the light curves are
undersampled in time, this may be treated as an upper limit only.

\titleb {3.3}{First Order Structure Function}

We calculated an average structure function over the sample by normalizing
the light curve of each source to its dispersion. This normalizes the
structure function at its asymptotic long time scale range and preserves
information on autocorrelation function over the averaging procedure,
assuming that the process is stationary. The dispersion is reasonably
approximated for the EGRET data since the light curves undersample the
variability time scales. On the other hand, an undersampled light curve of
pure stochastic stationary process should give flat structure function
(Rutman 1978) which was indeed found in some EGRET data (e.g., Wagner
1996; von Montigny \& Wagner 1996). The average structure function we
derived (Fig.~4) fails the standard behavior, and indicates clearly
internal correlations on time scales longer than $\sim$ 1200 days. 

\begdoublefig 0 cm
{\centerline{\epsfysize=6.3truecm\epsfbox{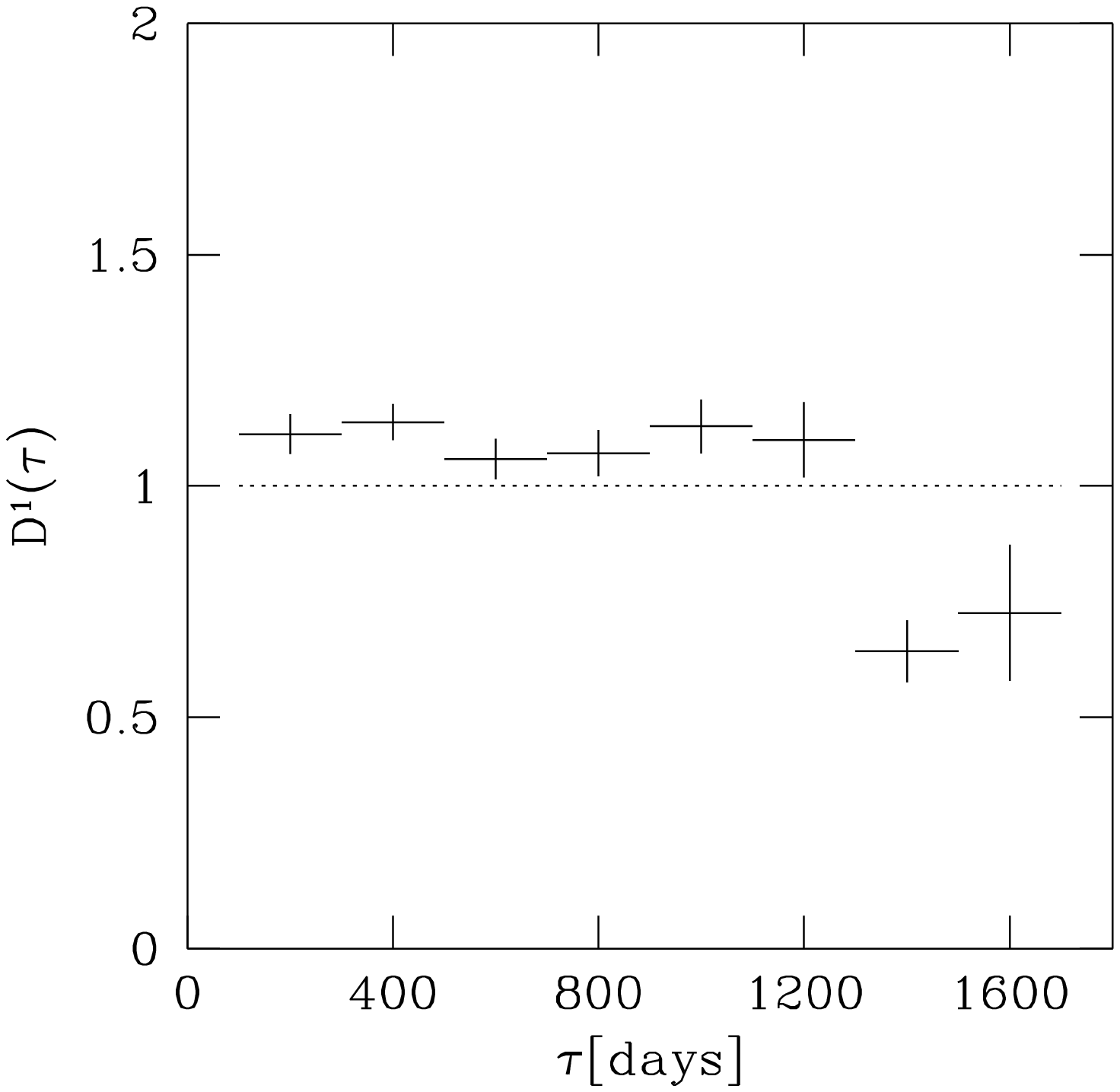}}
\figure{4}{
The average first order structure function in the sample of EGRET
blazars. The time process is normalized to its average dispersion.    
Vertical error bars indicate 95 per cent confidence limit.
}}
{\centerline{\epsfysize=6.3truecm\epsfbox{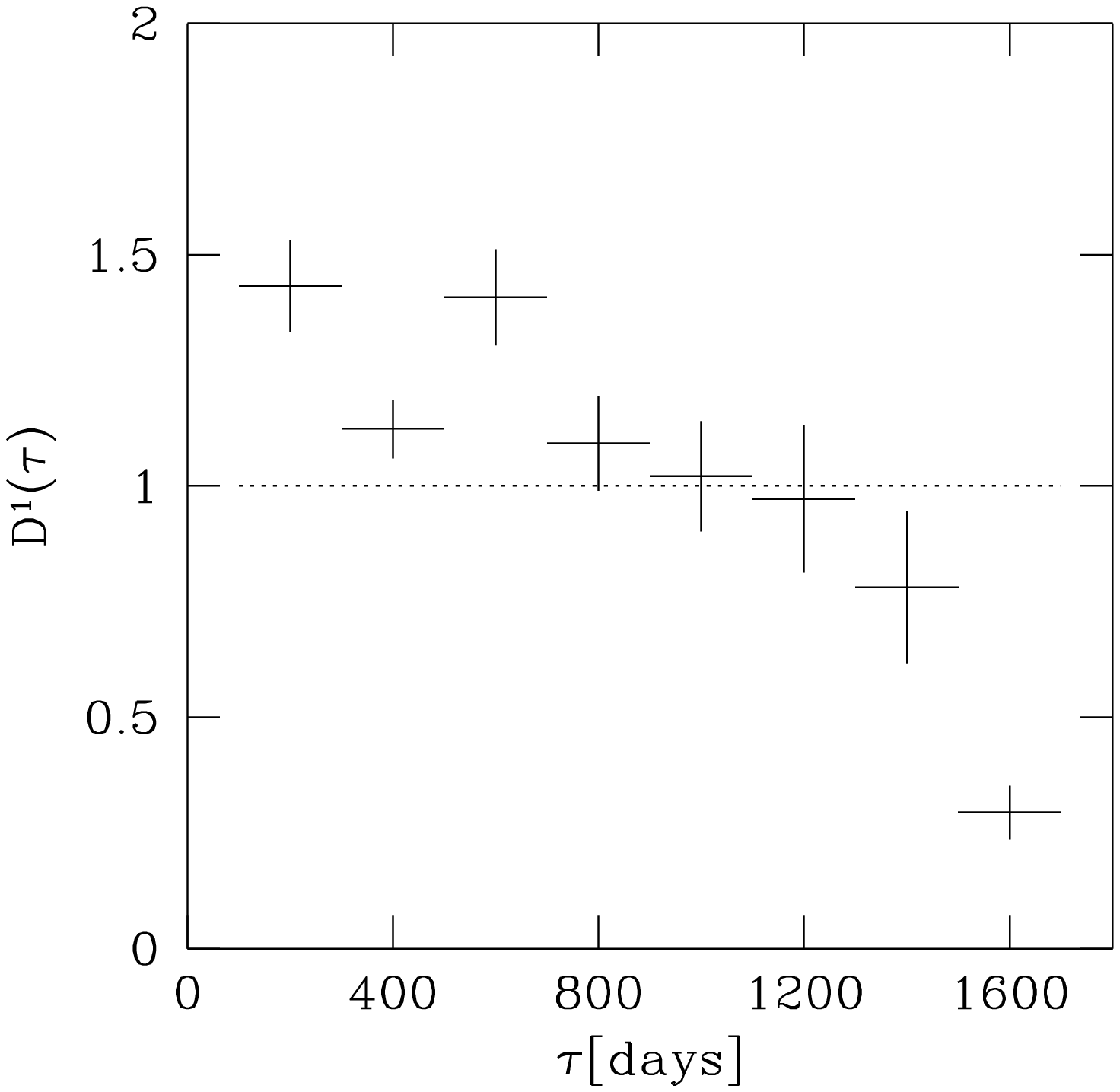}}
\figure{5}{
The average structure function of measurement errors, calculated under
assumption that the internal flux of each source is constant
(normalization is the same as on Fig.~4). 
}}
\enddoublefig

\noindent 

The structure function is additive with any uncorrelated short time scale
noise, well representing measurement errors (Simonetti et al.\ 1985). This
allows the determination of the basic level of observational artifacts
affecting the structure function. Fig.~5 presents the average structure
function calculated over the observations of EGRET sample, but under
assumption of constant internal source flux. The function, however, shows
similar basic level and similar behavior as that calculated for the
measurement flux. This indicates that the time process is dominated by
measurement errors or by the nature of the sampling, and the flux
measurements are indeed strongly correlated with their errors.

\titlea{4}{Conclusions}

Our preliminary analysis shows that the variability pattern in the EGRET
sample of blazars is strongly affected by measurement errors, and correct
understanding of the flux measurement statistics is crucial for
understanding physical content of the EGRET data. The variability shows
generally uniform pattern with power law distribution of temporal flux
changes. All of the statistics are consistent with non-linear models of
variability which may produce, contrary to the linear ones, flaring light
curves with saturation of time scales (e.g., Vio et al.\ 1992). There is a
little evidence in the structure function for suppressing the long time
scales which may be manifestation of reduced flaring probability near
large flares. This behavior is well described by self-organizing
critically models, and seems to be universal in accreting systems (Leighly
\& O'Brien 1997). 

\bigskip\noindent
{\it Acknowledgments} We thank Dr. R. C. Hartman for helpful discussion.

\begrefchapter{References}
\ref
Hartman, R. C.
1996, 1996, ASP Conf. Ser., Vol. 110,
Blazar Continuum Variability,
eds. Miller, H. R., Webb, J. R., \& Noble, J. C., p. 334
\ref
Leighly, K. M., \& O'Brien, P. T.
1997, ApJ, 481, L15
\ref
Lindsay, W. E., \& Chie, C. M. 
1976, Proc. IEEE, 64, 1652
\ref
Madejski, G. M., et al.
1997, {\it X-ray Imaging and Spectroscopy of Cosmic Hot Plasmas},
eds. Makino, F., \& Mitsuda, K., p. 229
\ref
Magdziarz, P., \& Machalski, J.
1993, A\&A, 275, 405
\ref
Mattox, J. R., et al.
1996, ApJ, 461, 396
\ref
McHardy, I.
1996, 1996, ASP Conf. Ser., Vol. 110,
Blazar Continuum Variability,
eds. Miller, H. R., Webb, J. R., \& Noble, J. C., p. 293
\ref
McLaughlin, M. A., et al.
1996, ApJ, 473, 763
\ref
Rutman, J.
1978, Proc. IEEE, 66, 1048
\ref
Simonetti, J. H., Cordes, J. M., \& Heeschen, D. S.
1985, ApJ, 296, 46
\ref
Sugihara, G., \& May, R.
1990, Nat, 344, 734
\ref
Thomson, D. J., et al.
1993, ApJS, 86, 629
\ref
Thomson, D. J., et al.
1995, ApJS, 101, 259
\ref
Urry, C. M.
1996, ASP Conf. Ser., Vol. 110, 
Blazar Continuum Variability, 
eds. Miller, H. R., Webb, J. R., \& Noble, J. C., p. 391
\ref
Urry, C. M., \& Padovani, P.
1995, PASP, 107, 803
\ref
Vio, R., et al.
1992, ApJ, 391, 518
\ref
von Montigny, C., \& Wagner, S.
1996, MPI H-V37-1996, 
Proc. of the Heidelberg Workshop on $\gamma$--ray emitting AGN,
eds. Kirk, J., et al., p. 113
\ref
Wagner, S.
1996, MPI H-V37-1996,
Proc. of the Heidelberg Workshop on $\gamma$--ray emitting AGN,
eds. Kirk, J., et al., p. 117
\ref
Willis, T. D.
1996, Ph.D. Thesis, Stanford University
\endref

\byebye